\documentclass{moriond}

\def\Journal#1#2#3#4{{#1} {\bf #2}, #3 (#4)}

\def\PRD{{\em Phys. Rev.} D}

\def\JCAP{\em JCAP}

\def\be{\begin{equation}}
\def\ee{\end{equation}}
\def\bea{\begin{eqnarray}}
\def\eea{\end{eqnarray}}

\newcommand{\Photo}{}
\usepackage{lineno}
\usepackage{minipage}
\usepackage{wrapfig}
\begin{document}
\vspace*{4cm}
\title{Recent IceCube Results: Diffuse Flux, Point Sources and Dark Matter}

\author{Minjin Jeong (on behalf of the IceCube Collaboration) }

\address{Department of Physics and Astronomy, University of Utah, USA}

\maketitle\abstracts{
In 2013, the IceCube Collaboration reported the first observation of an astrophysical neutrino flux, with energies extending up to the PeV-scale. Over the last decade, this flux has been characterized by measurements in multiple detection channels that are complementary with respect to the sensitive energy range, flavor composition, sky coverage, and backgrounds. The origin of these neutrinos remains largely unknown. However, evidence has been found for neutrino emission from the directions of the blazar TXS 0506+056, Seyfert galaxy NGC 1068. and the Galactic Plane. IceCube also has an active program of indirect dark matter searches with competitive constraints on dark matter models. In this talk, we will present recent results of the IceCube experiment, highlighting the latest diffuse flux measurements, point source searches, and dark matter analyses.}

\section{Introduction}
IceCube~\cite{detector} is a neutrino detector located at the geographic South Pole. The detector consists of 5,160 digital optical modules (DOMs), each containing a photomultiplier tube (PMT). These DOMs are distributed throughout a cubic-kilometer volume of exceptionally transparent glacial ice at depths between 1,450~m and 2,450~m. The DOMs can detect the Cherenkov radiation associated with relativistic charged particles produced when neutrinos interact in the ice. In the bottom-center of IceCube, they are more densely spaced than in the rest of the detector. This region, referred to as the DeepCore subarray, was designed to lower the energy threshold of IceCube by over an order of magnitude, reaching as low as about 10 GeV~\cite{DeepCore}. 

IceCube events are usually classified into two types: tracks and cascades. Tracks are predominantly generated by muon neutrinos undergoing a charged-current (CC) interaction in the ice. In general, these events provide excellent angular resolution, although their energy reconstruction is challenging. Cascades are produced by electron and tau neutrinos undergoing a CC interaction or by a neutral-current (NC) interaction of all neutrino flavors. While cascades offer superior energy resolution compared to tracks, their directional reconstruction is more challenging. A rare event topology, known as double cascades or double bangs, can occur from a CC interaction of high-energy tau neutrinos. This interaction produces a tau lepton which traverses the ice and can decay into a tau neutrino and a hadronic shower. This hadronic shower can then generate a second cascade. 

In 2013, IceCube initiated a new era of neutrino astronomy by observing a diffuse astrophysical neutrino flux~\cite{science_2013_diffuse}. Since then, the diffuse astrophysical flux has been characterized through various detection channels. Although the origin of these neutrinos remains largely unknown, IceCube has found evidence for neutrino emissions from the directions of TXS 0506+056, NGC 1068, and the Galactic Plane~\cite{TXS_1st,TXS_2nd,science_2022_NGC1068,galactic_plane}. Furthermore, some of our latest analyses hint at potential event excesses associated with NGC 4151 and CGCG 420-015~\cite{ICRC2023_hard-X-ray_AGNs,ICRC2023_Seyfert_Northern_Sky}. IceCube also conducted dark matter (DM) searches for over a decade, considering various targets, DM masses and interaction channels, and has derived highly competitive limits on the DM parameter space~\cite{Galactic_DM_2011,Galactic_DM_2015,Galactic_DM_2018,Galactic_DM_2020,solar_DM_2017,Earth_DM,ICRC2021_Secluded_DM}. In this contribution, we present recent diffuse flux measurements, point source searches, and DM searches with IceCube. 

\section{Measurements of Diffuse Neutrino Flux}

The existence of an astrophysical neutrino flux was first established in 2013 by analyzing two years of High Energy Starting Events (HESE). The HESE data include tracks and cascades that start inside the detector, covering the entire sky, with an energy threshold of 60~TeV. Subsequent iterations of the HESE analysis, with increased sample sizes, observed an astrophysical neutrino flux consistent with an isotropic flux following a single power-law (SPL) spectrum model~\cite{science_2013_diffuse}. The astrophysical flux has also been measured through other detection channels, such as through-going tracks from the Northern Sky~\cite{diffuse_NT}, cascades~\cite{diffuse_cascades}, the Medium Energy Starting Events (MESE)~\cite{MESE_2-year}, and a special event sample established for inelasticity measurements~\cite{diffuse_inelasticity}. These different channels are complementary with respect to their sensitive energy range, sky coverage, and neutrino flavors. These measurements have been consistent with the SPL hypothesis. Although hints for substructures in the spectrum were seen from independent measurements, they were statistically insignificant~\cite{diffuse_NT,diffuse_cascades,MESE_2-year}.       

\begin{figure}[tb]
\centering
    \includegraphics[width=0.37\linewidth]{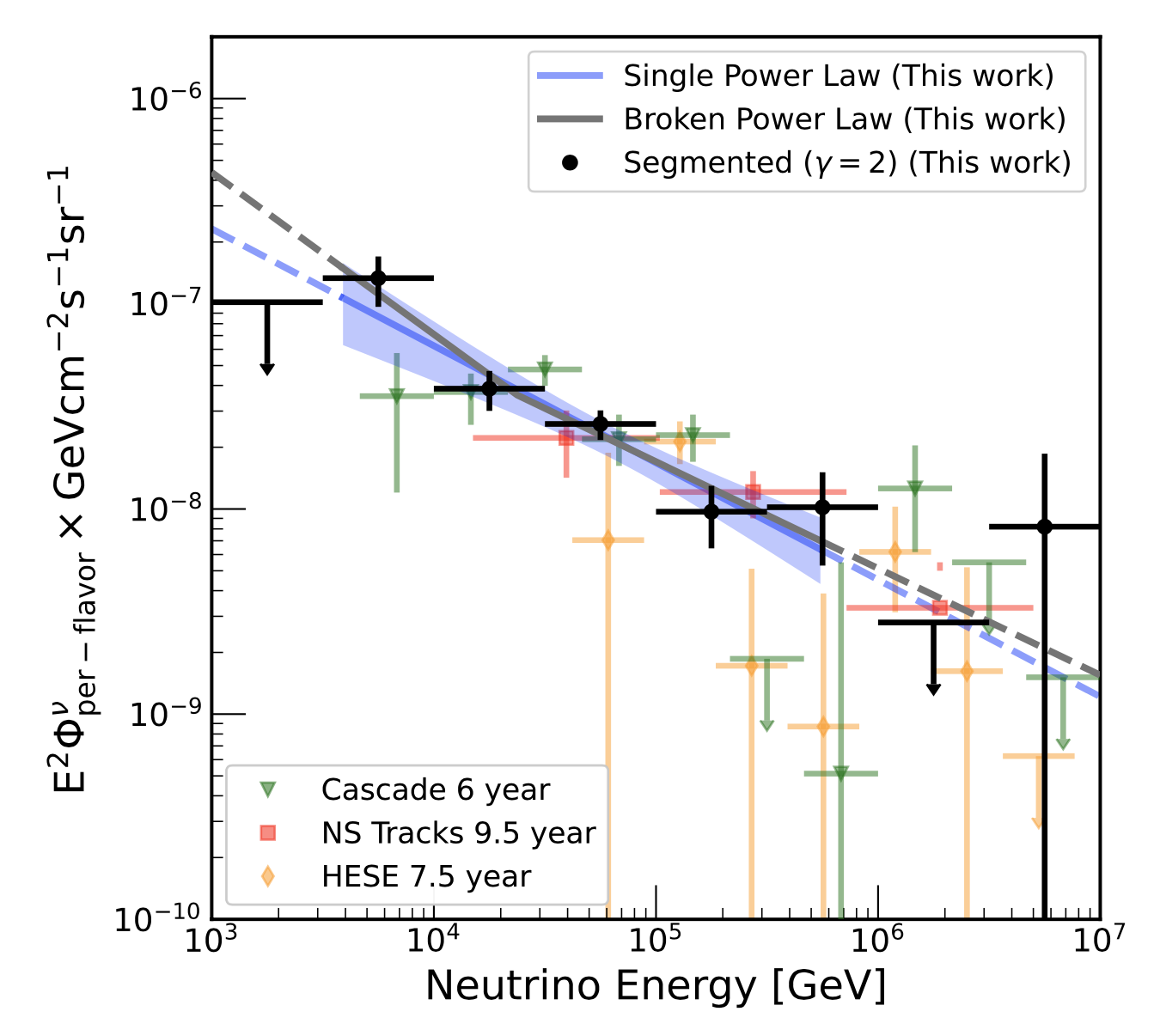}
    \includegraphics[width=0.44\linewidth]{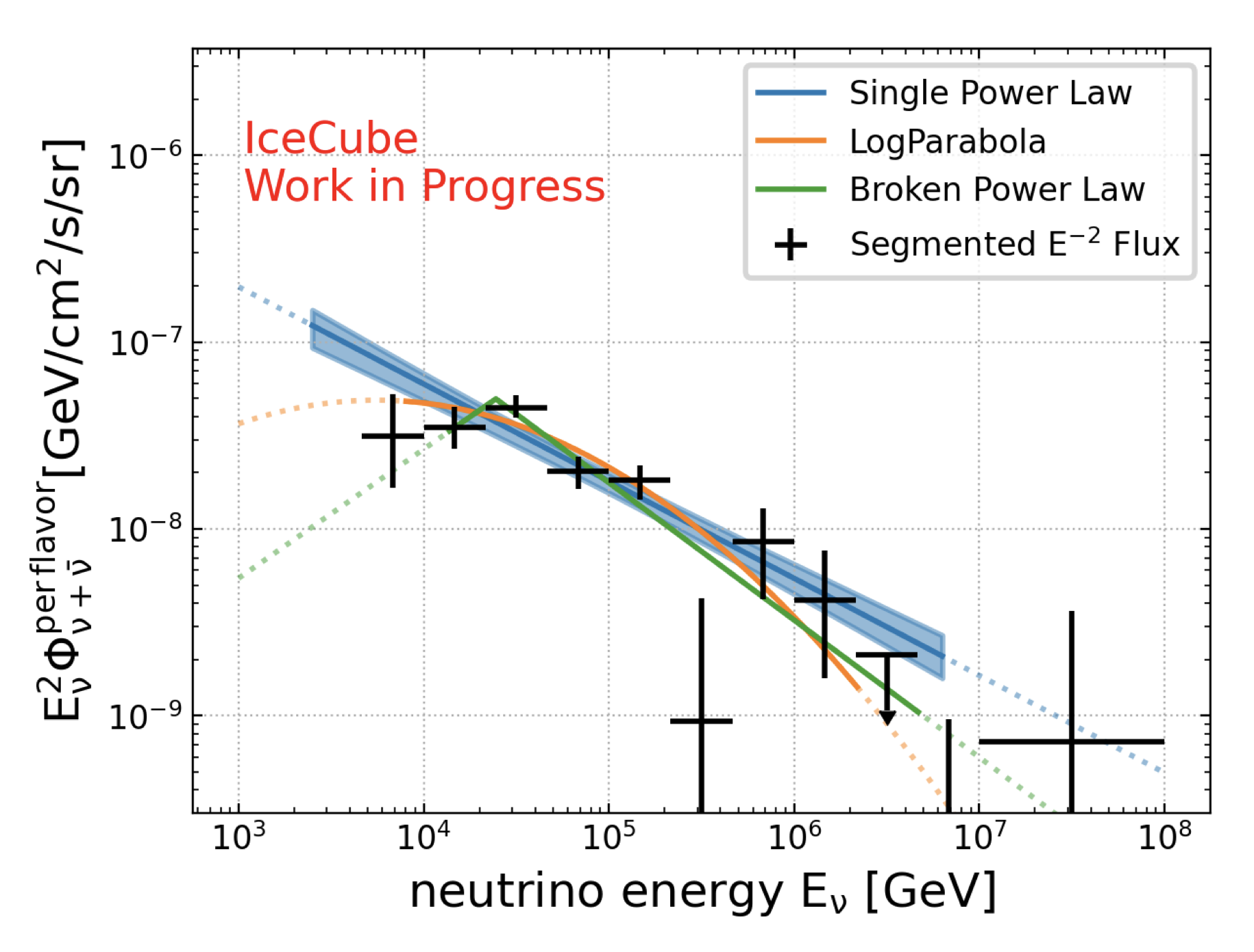}
    \caption{Diffuse astrophysical neutrino spectrum measured with the ESTES analysis (left panel) and combined fit of tracks and cascades (right panel). In the left panel, the ESTES results are compared with the previous IceCube measurements with different detection channels. These plots are taken from References ~\protect\cite{ESTES_paper,ICRC2023_global-fit}.}
    \label{fig:ESTES_and_global-fit_spectrum}
\end{figure}

A new detection channel established recently is the Enhanced Starting Tracks Event Selection (ESTES)~\cite{ESTES_paper}. ESTES is designed to select tracks that have interaction vertices contained in the detector. The event selection utilizes a dynamic veto as well as Boosted Decision Trees to reduce atmospheric muon backgrounds, and achieves neutrino purity of 99.3\%. The track topology provides an excellent median angular resolution, which is 1.6 degrees at 1~TeV neutrino energy and 0.66 degrees at 100~TeV. Moreover, the hadronic shower at the event vertex leads to an energy resolution of 25-30\% for neutrino energies between 1~TeV and 10~PeV, which is superior to through-going track samples. We characterized the astrophysical neutrino flux using this sample, assuming an isotropic flux and several spectrum models. These results are compared with previous astrophysical neutrino spectrum measurements in the left panel of Figure~\ref{fig:ESTES_and_global-fit_spectrum}. The ESTES results are consistent with the previous IceCube measurements and cover lower energies than the previous analyses, down to 3~TeV.  

\begin{figure}[bt]
    \centering
    \includegraphics[width=2.7in]{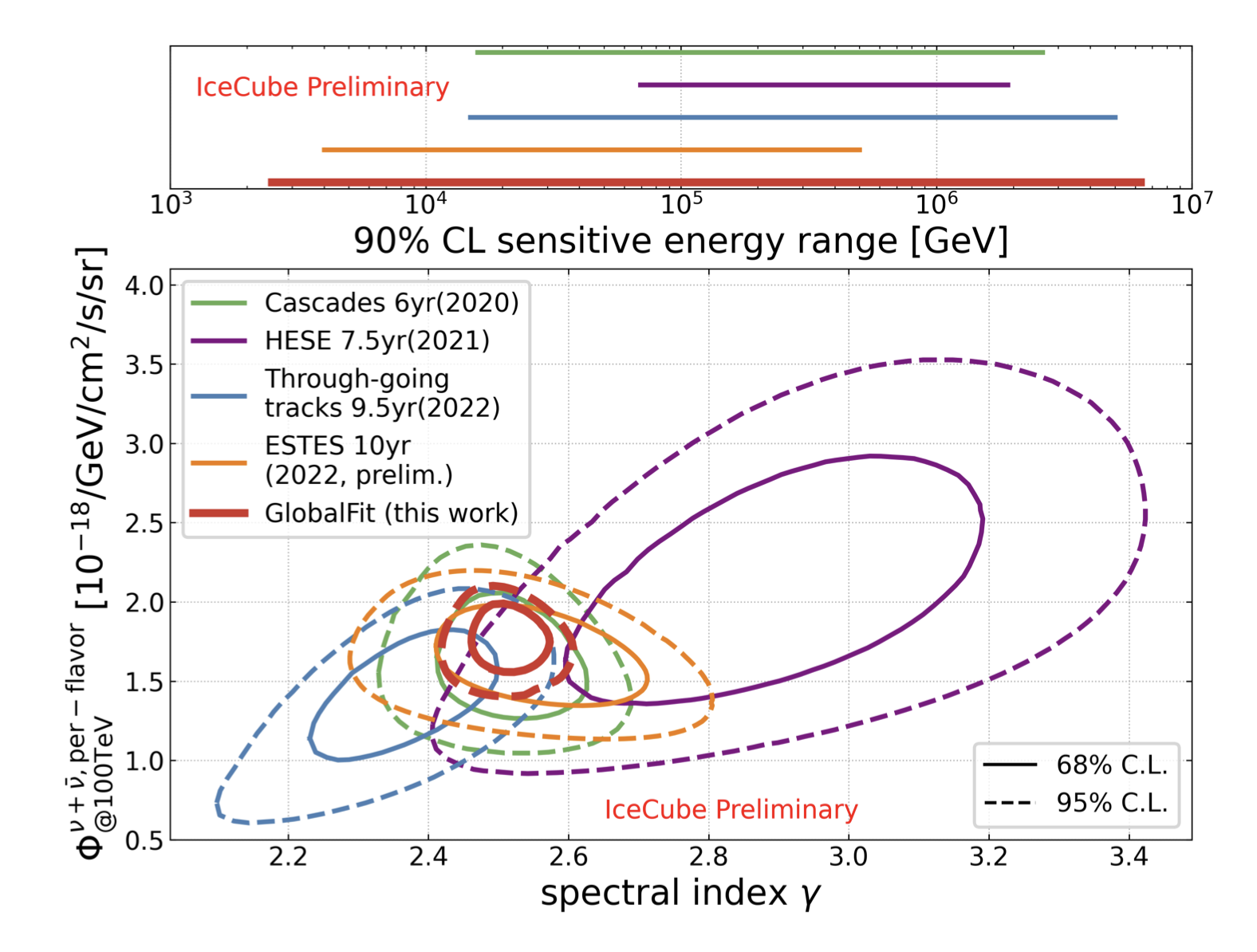}
    \includegraphics[width=2.5in]{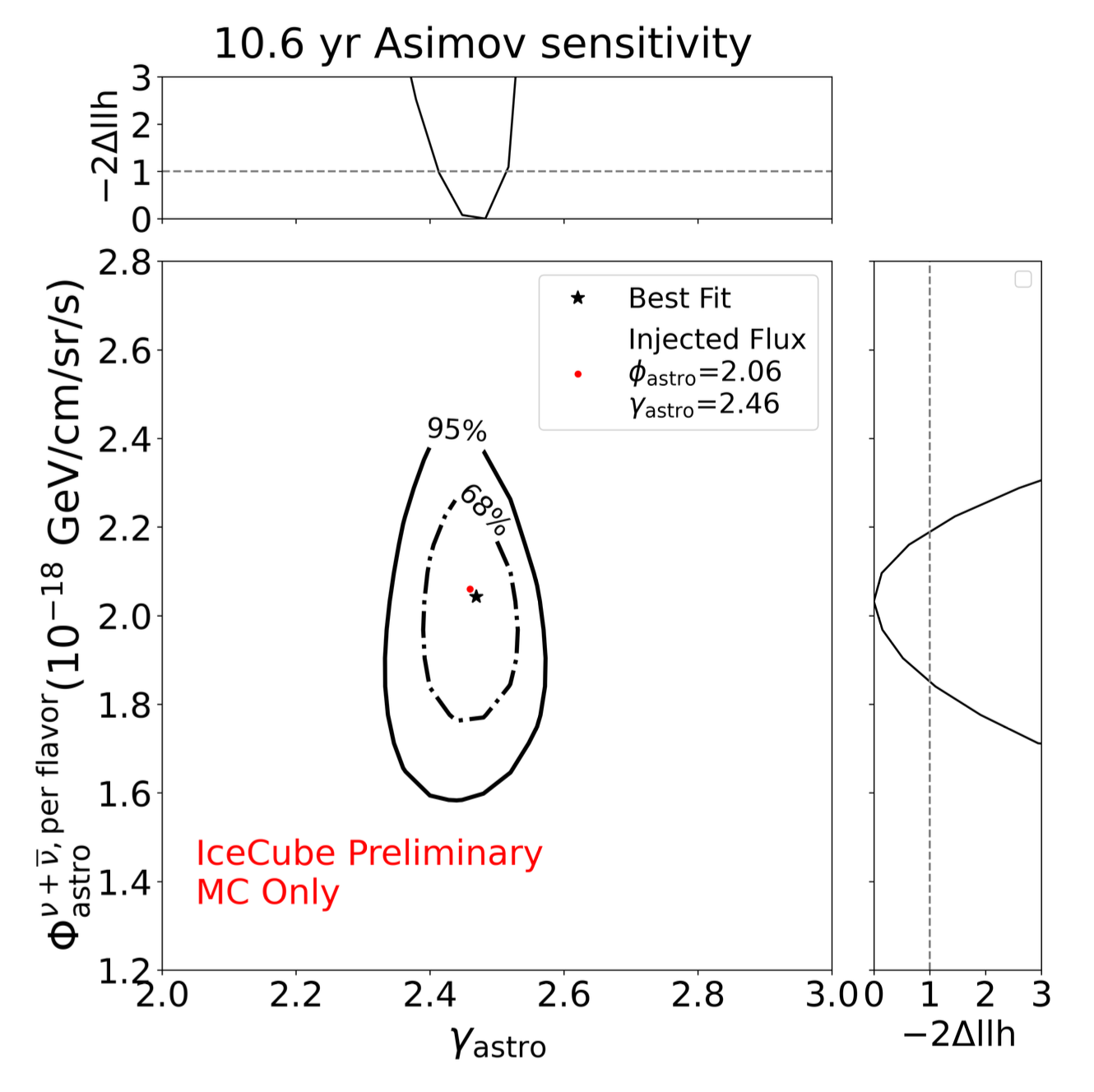}
    \caption{The left panel shows the pectral index and normalization for diffuse astrophysical neutrino flux measured under a SPL spectrum hypothesis. The right panel shows the sensitivity of the MESE 10.6-year analysis to the normalization and spectral index. To calculate the sensitivity, a flux was injected with the best-fit normalization and spectral index reported from the MESE 2-year analysis~\protect\cite{MESE_2-year}. These plots are taken from References~\protect\cite{ICRC2023_global-fit,ICRC2023_MESE}.}
    \label{fig:combined_fit_and_MESE_spectrum}
\end{figure}

Combining different detection channels is critical for accurately characterizing the diffuse astrophysical neutrino flux. The latest global-fit analysis~\cite{ICRC2023_global-fit} combines 8.5 years of through-going tracks that originate from the Northern Sky and 10.5 years of all-sky cascades. We treated nuisance parameters that model the atmospheric flux uncertainties and detector systematic parameters in a consistent way across the two channels. As shown in the right panel of Figure~\ref{fig:ESTES_and_global-fit_spectrum}, the measured astrophysical neutrino spectrum is consistent with a SPL spectrum hypothesis, while more complex models such as a broken power-law model are favored. The spectral index and normalization estimated under the SPL hypothesis are shown in Figure~\ref{fig:combined_fit_and_MESE_spectrum}. The estimates are consistent with previous IceCube measurements. Moreover, this global-fit has achieved unprecedented precision. 

The MESE selection extends the HESE selection down to a neutrino energy of 1~TeV, while retaining high signal purity. A previous analysis using 2 years of MESE data showed a possible event excess above atmospheric backgrounds at neutrino energies around 30~TeV~\cite{MESE_2-year}. Our latest MESE analysis, which utilizes 10.6 years of data, aims to investigate the astrophysical neutrino spectrum with the increased sample size as well as improved understanding of the atmospheric backgrounds~\cite{ICRC2023_MESE}. The right panel of Figure~\ref{fig:combined_fit_and_MESE_spectrum} shows the expected sensitivity of the analysis to the astrophysical neutrino spectrum. The analysis results will be presented in our future publications.

\section{Searches for Neutrino Point Sources}
IceCube observed evidence for neutrino emission in the direction of NGC 1068 with a post-trial significance of 4.2 $\sigma$~\cite{science_2022_NGC1068}, reinforcing the idea that Active Galactic Nuclei (AGN) are high-energy cosmic ray accelerators. The observed neutrino flux at the TeV-scale is an order of magnitude greater than the upper limit for the gamma-ray flux~\cite{MAGIC_limit_NGC1068,HAWC_limit_NGC1068}. Furthermore, previous IceCube searches targeting gamma-ray bright AGNs did not reveal any significant excess of events~\cite{blazar_2017,ICRC2019_blazar}. Consequently, theoretical models have suggested that high-energy neutrinos might be produced in regions opaque to gamma rays~\cite{AGN_theory,Disk_corona_1,Disk_corona_2}. In this contribution, we present our recent point source searches that aimed to test this hypothesis.

\begin{figure}[tb]
    \centering
    \includegraphics[width=0.7\linewidth]{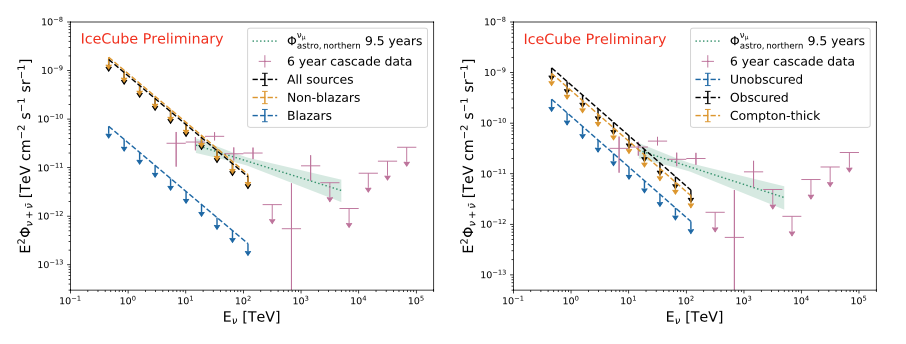}
    \caption{Upper limits on the collective neutrino emission from AGNs bright in the hard X-ray band. The left panel shows the limits for all AGNs, blazars, and non-blazars. The right panel presents the limits for unobscured AGNs, obscured, and Compton-thick AGNs. In both panels, the limits are compared with the diffuse astrophysical flux measured by previous IceCube analyses. These plots are from Reference~\protect\cite{ICRC2023_hard-X-ray_AGNs}.}
\label{fig:limits_X-ray_AGNs}
\end{figure}

In a recent study, a 12-year all-sky track sample was analyzed to search for neutrinos from AGNs that are bright in the hard X-ray band (14-195 keV)~\cite{ICRC2023_hard-X-ray_AGNs}. For this work, we selected 836 AGNs from the Swift-BAT Spectroscopic Survey (BASS) catalog, which covers the entire sky and is, to date, the most complete list of AGNs detected in the hard X-ray band. Our stacking search led to upper limits on the cumulative neutrino emission from all AGNs and the following subclasses: blazars, non-blazars, unobscured sources, obscured sources, and Compton-thick sources.  Figure~\ref{fig:limits_X-ray_AGNs} displays the limits. We found that blazars contribute at most 7\% to the total diffuse neutrino flux measured using muon tracks at 100~TeV. Additionally, we searched for signals from each of the 43 most promising sources individually and observed an event excess from the direction of NGC 4151 at a post-trial significance of 2.9~$\sigma$. 

\begin{figure}[tb]
    \centering
    \includegraphics[width=0.38\linewidth]{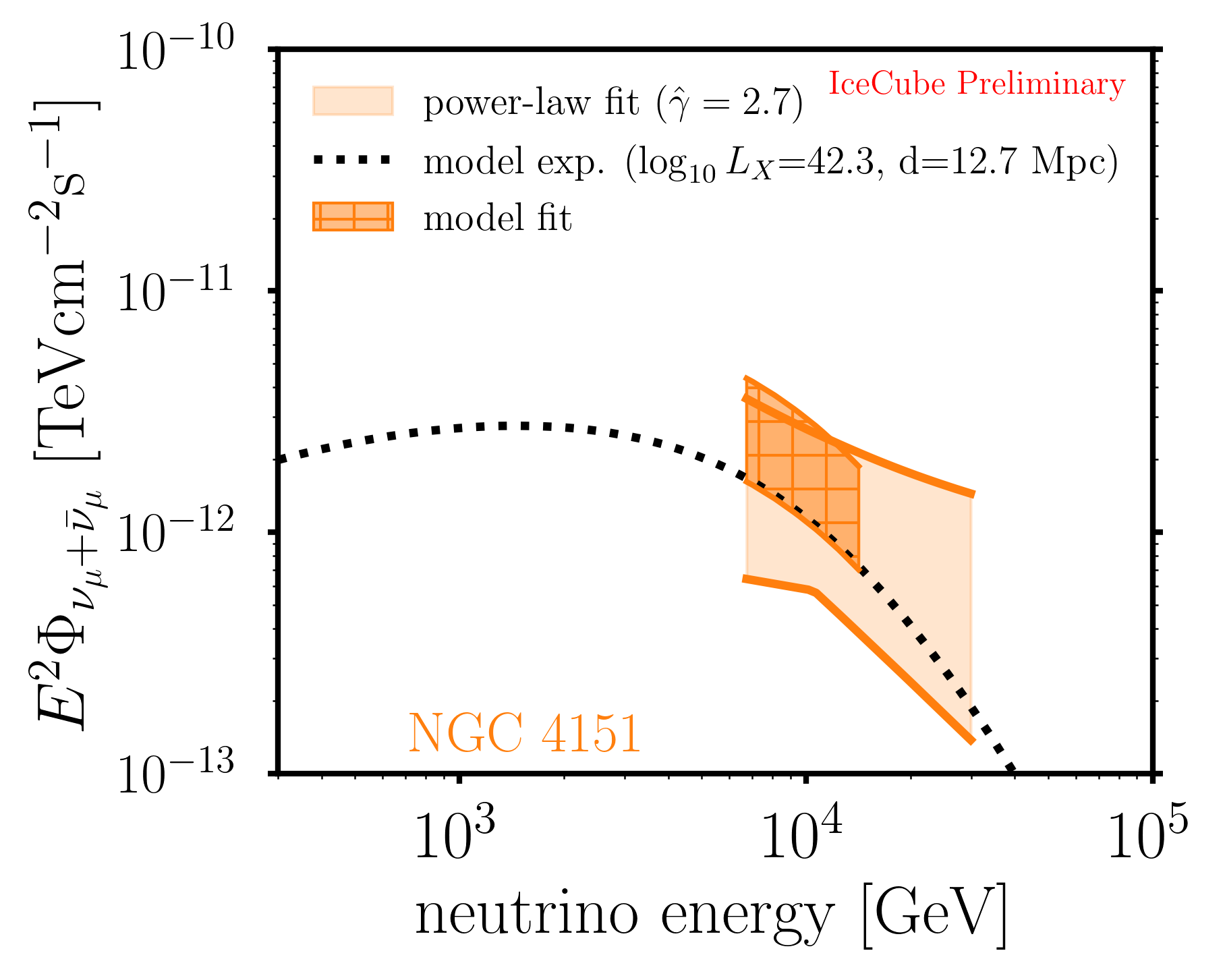}
    \includegraphics[width=0.38\linewidth]{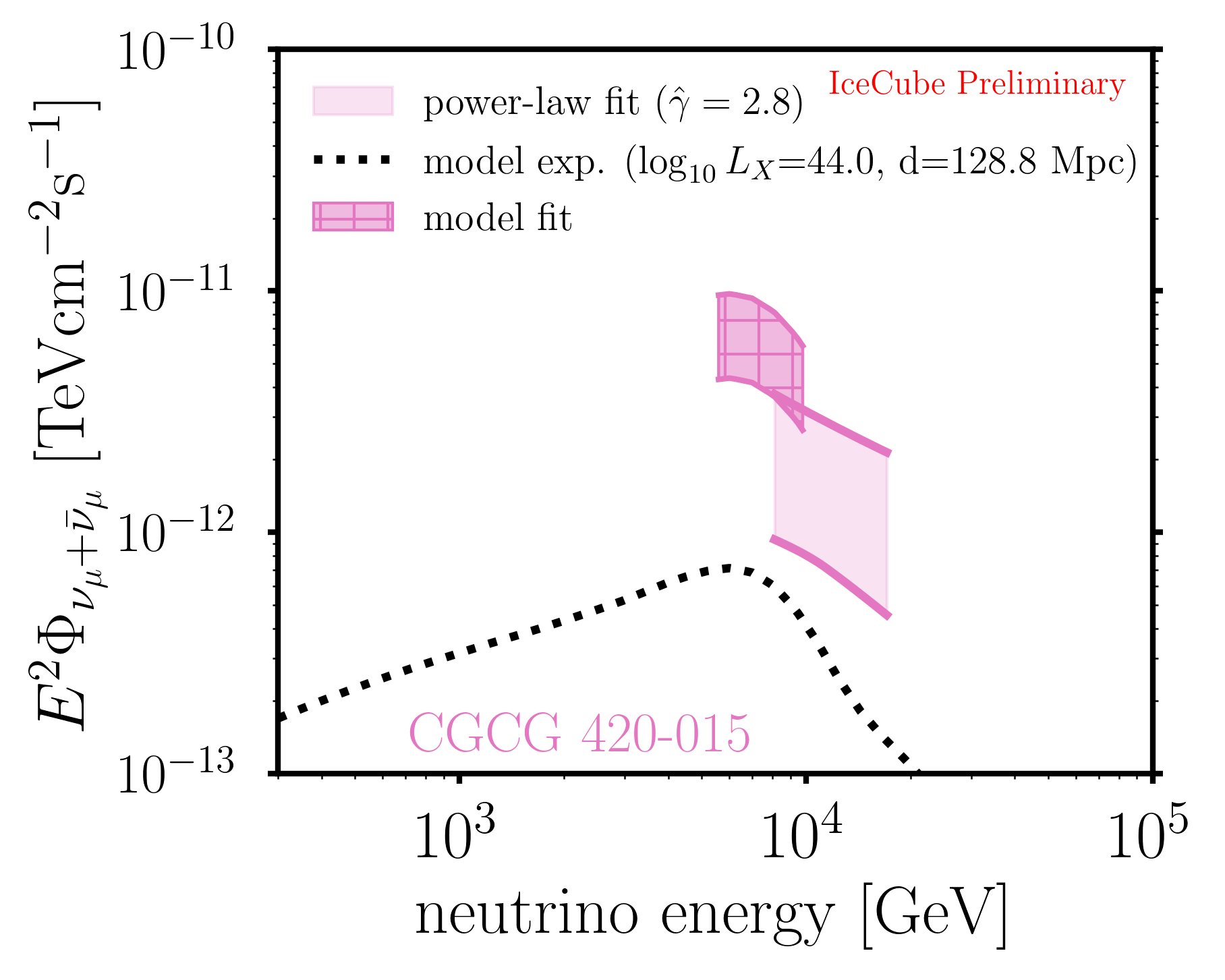}
    \caption{Neutrino fluxes fitted in the direction of NGC 4151 and CGCG 420-015 in the Northern Sky search for Seyfert galaxies. These plots were presented in Reference~\protect\cite{TAUP-2023}.}
    \label{fig:Results_of_Northern_Sky_Seyfert_analysis}
\end{figure}

An independent analysis investigated neutrino emission from X-ray bright Seyfert galaxies in the Northern Sky~\cite{ICRC2023_Seyfert_Northern_Sky}. For this study, we selected 28 sources with high intrinsic X-ray brightness from the BASS catalog and analyzed 10.4 years of through-going tracks from the Northern Sky. We searched for signals from the individual sources considering neutrino spectra predicted by the disk-corona model~\cite{Disk_corona_2} as well as the SPL model, and found a 2.7~$\sigma$ event excess associated with CGCG 420-015 and NGC 4151. The fluxes fitted for the two sources are shown in Figure~\ref{fig:Results_of_Northern_Sky_Seyfert_analysis}. Additionally, we performed a stacking search using the selected sources except NGC 1068 and constrained the collective neutrino emission from those Seyfert galaxies. 

\begin{figure}[tb]
\centering
\includegraphics[width=0.36\linewidth]{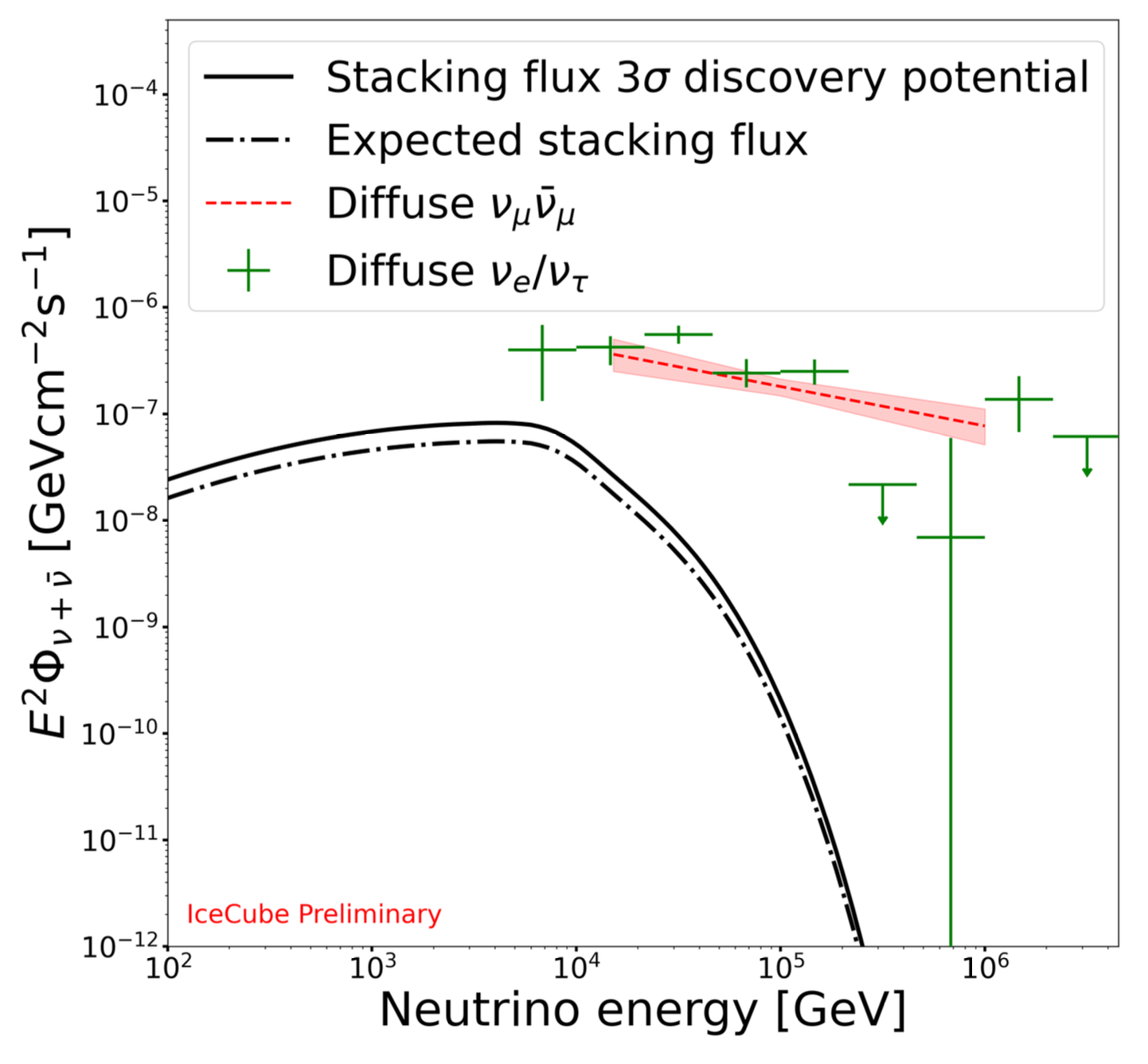}
\caption{Discovery potential of the collective neutrino emission from Seyfert galaxies in the Southern Sky. The solid line represents the discovery potential calculated at the 3~$\sigma$ level. The dash-dotted line indicates the predicted neutrino flux according to the disk-corona model. These are compared with previous diffuse astrophysical flux measurements. This figure is taken from Reference~\protect\cite{ICRC2023_Seyfert_Southern_Sky}.}
\label{fig:Discovery_potential_Southern_Sky_Seyfert}
\end{figure}

A similar study is in progress targeting Seyfert galaxies in the Southern Sky~\cite{ICRC2023_Seyfert_Southern_Sky}. Fourteen sources were selected from the BASS catalog accounting for their intrinsic X-ray brightness and distance from Earth. We are using the ESTES 10.3-year sample, and considering the disk-corona model and SPL model to estimate the neutrino spectrum. Similar to the Northern Sky analysis, we plan to search for signals from the individual sources as well as for collective neutrino emissions from the sources. For the stacking search, we exclude Centaurus A, as its high X-ray brightness could originate from a jet rather than its corona. Figure~\ref{fig:Discovery_potential_Southern_Sky_Seyfert} presents the 3~$\sigma$ discovery potential of the stacking search and the signal flux predicted using the disk-corona model. The analysis results will be discussed in our future publications.

\section{Indirect Searches for Dark Matter}
A primary technique for detecting DM is based on the assumption that DM particles can self-annihilate or decay into Standard Model particles, resulting in messenger particles, such as neutrinos, in the final state. In this scenario, a flux of neutrinos is expected from massive celestial objects like galaxy clusters, galaxies, and the Galactic Center. The directional distribution of the flux is associated with the DM distribution in the source, and the spectrum depends on the primary products of the DM interaction. This flux can be very different from the atmospheric and astrophysical neutrino backgrounds, depending on the DM mass, interaction channel, and the DM distribution. Therefore, neutrino detectors like IceCube can test DM hypotheses by looking at celestial objects that are rich in DM. This technique is referred to as an indirect DM search. 

\begin{figure}[tb]
\centering
    \includegraphics[width=0.9\linewidth]{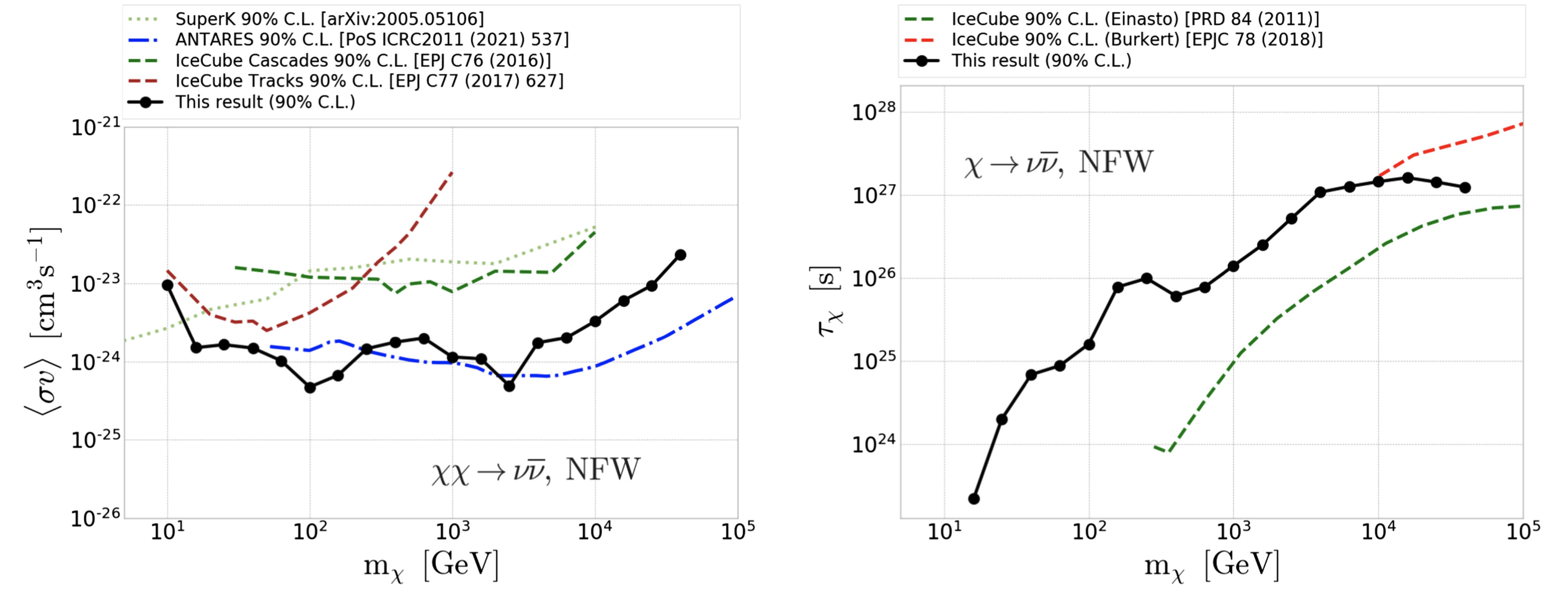}
    \caption{Limits from the neutrino line analysis on the DM self-annihilation cross section (left panel) and lifetime (right panel). In both panels, the solid lines are the limits derived by the neutrino line analysis, and the others by previous neutrino experiments. All limits here were calculated assuming that the Galactic DM distribution follows a Navarro-Frenk-White (NFW) profile. These plots are taken from Reference~\protect\cite{NuLine}.  }
    \label{fig:NuLine_limits}
\end{figure}

The Galactic Center and Halo are some of the most promising targets for indirect DM searches, due to their proximity and high DM content. Hence, IceCube looked for signals from Galactic DM annihilation and decay for a long time. In our recent study~\cite{NuLine}, we considered DM masses ranging from 10~GeV to 40~TeV and optimized the analysis methods to detect signals from the $\chi \rightarrow \nu\bar{\nu}$ and $\chi\chi \rightarrow \nu\bar{\nu}$ scenarios. In these scenarios, the expected neutrino spectrum includes a line, and its observation would constitute a smoking gun signal of DM. We selected 5 years of cascade events contained in the DeepCore subarray specifically for this analysis, achieving an energy resolution of 30\% for neutrino energies above 100~GeV. The analysis derived significantly improved limits upon previous IceCube searches for Galactic DM annihilation and decay, as shown in Figure~\ref{fig:NuLine_limits}.

\begin{figure}[tb]
\centering
    \includegraphics[width=0.44\linewidth]{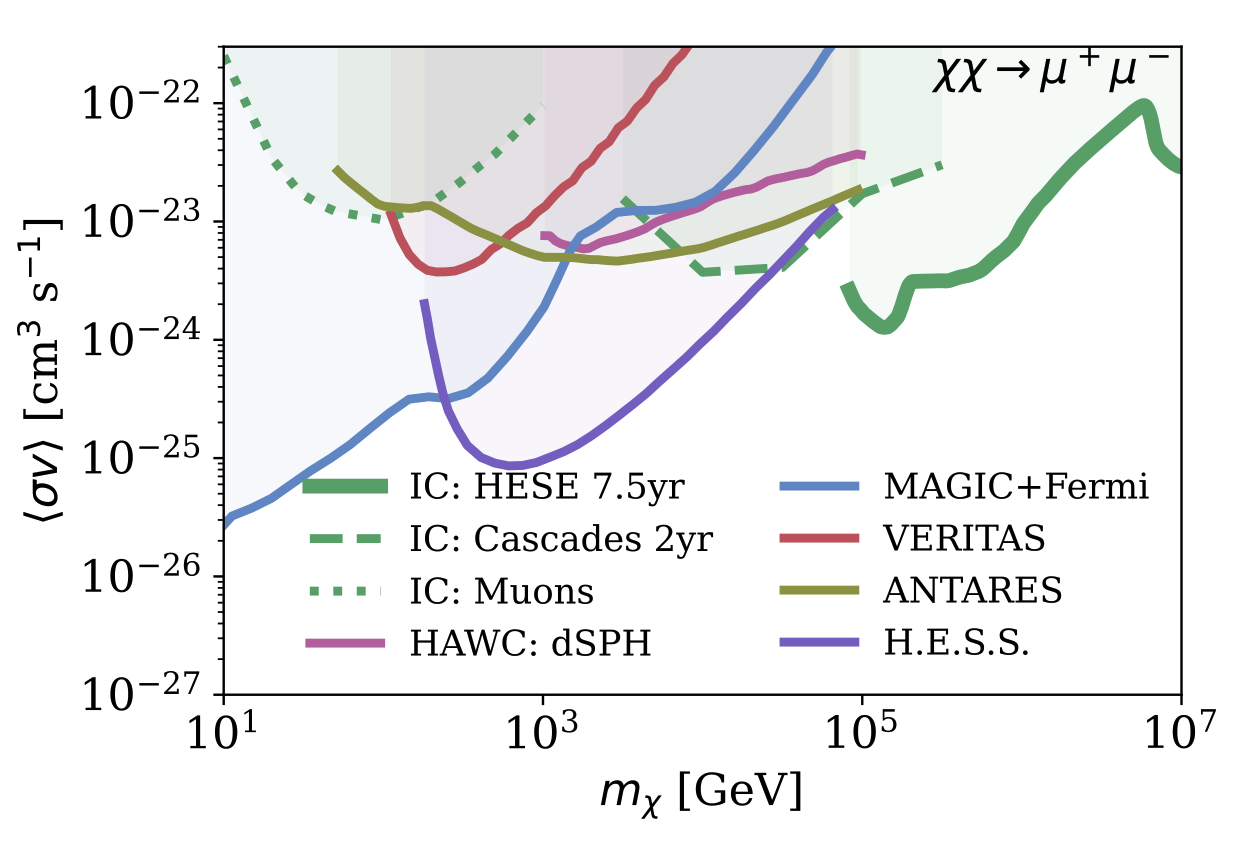}
    \includegraphics[width=0.46\linewidth]{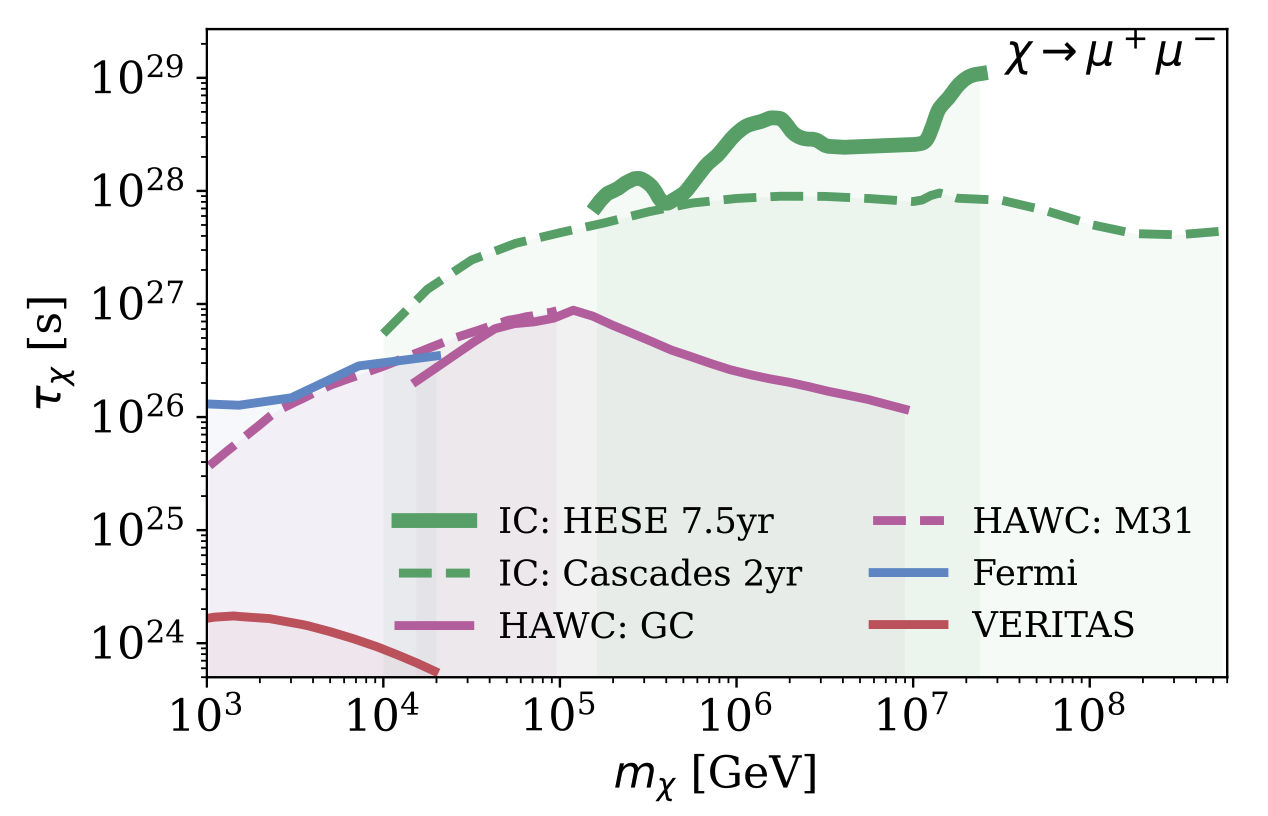}

    \caption{Limits from the HESE 7.5-year analysis on the DM self-annihilation cross section (left panel) and lifetime (right panel). In both panels, the thick solid lines were obtained by the HESE 7.5-year analysis, and the other lines by previous neutrino and gamma-ray experiments. The HESE limits were calculated assuming that the Galactic DM distribution follows an Einasto profile. These plots are taken from Reference~\protect\cite{HESE_7.5-year_DM}. }
    \label{fig:HESE_limits}
\end{figure}

DM masses on the GeV-to-TeV scale are particularly interesting, since the range is permissible for Weakly Interacting Massive Particles (WIMPs), which are among the most popular DM particle candidates. Thus, many traditional IceCube analyses have focused on probing this mass range. However, interest in alternative DM candidates such as heavy DM is growing. One reason for this is that WIMPs have not been detected despite numerous experimental efforts. Furthermore, modern instrumentation techniques allow us to test DM scenarios beyond the typical WIMP masses. In particular, the potential event excesses observed in some diffuse neutrino spectrum measurements have led to speculations about signals from PeV-scale DM~\cite{HESE_excess_DM_Esmaili13,superheavy_DM,MESE_excess_DM}. To test the heavy DM models, we searched 7.5 years of HESE data for neutrinos from Galactic and cosmological DM interactions~\cite{HESE_7.5-year_DM}. This analysis considers both annihilating DM, with masses ranging from 80~TeV to 10~PeV, and decaying DM, from 160~TeV to 20~PeV. Figure~\ref{fig:HESE_limits} shows limits obtained from this analysis.


\begin{figure}[tb]
\centering
    \includegraphics[width=0.64\linewidth]{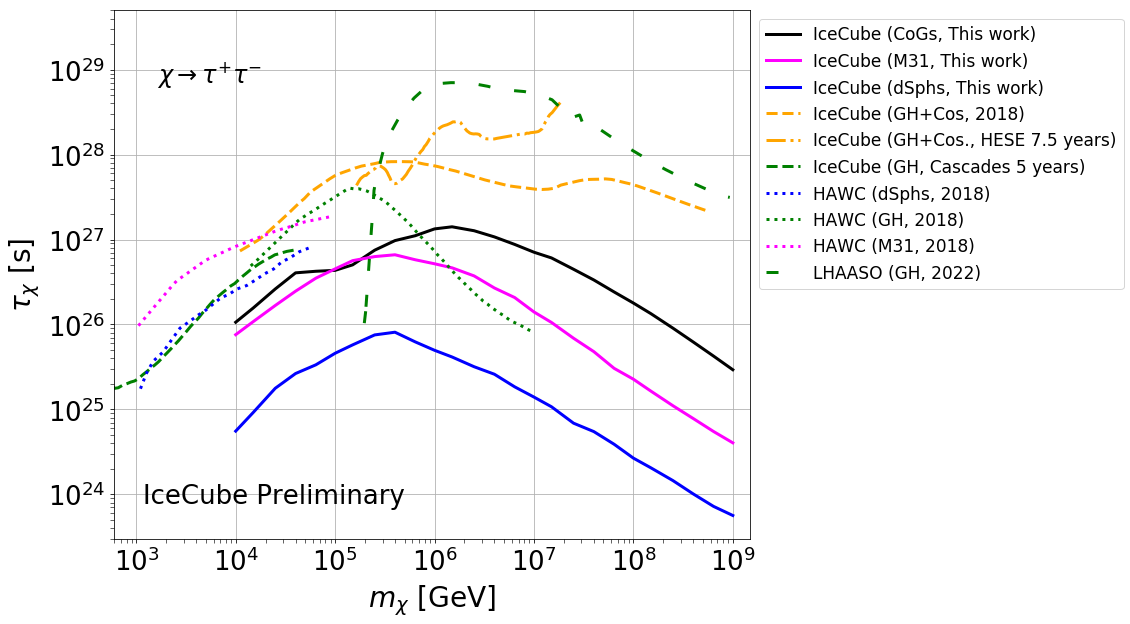}
    \caption{Limits from the extragalactic DM decay analysis on the DM lifetime. The solid lines represent the limits derived by IceCube. The other lines are the limits from previous DM searches with IceCube, HAWC, and LHAASO. The line colors indicate different targets used for the analyses: clusters of galaxies (black), dwarf spheroidal galaxies (blue), the Andromeda galaxy (magenta), the Galactic DM (green), and a combination of the Galactic and cosmological DM (orange). This figure is taken from Reference~\protect\cite{ICRC2023_extragalactic_DM_decay}. }
    \label{fig:ExtGal_DM_limits}
\end{figure}

The latest analysis searched for neutrinos from decaying heavy DM in nearby galaxy clusters and galaxies for the first time~\cite{ICRC2023_extragalactic_DM_decay}. This analysis focuses on DM masses ranging from 10~TeV to 1~EeV. Three galaxy clusters, the Andromeda galaxy, and seven dwarf spheroidal galaxies in the Northern Sky were chosen as targets, accounting for their predicted signal strengths. We stacked the galaxy clusters and dwarf galaxies within the same source class and utilized 10.4 years of through-going tracks from the Northern Sky. Figure~\ref{fig:ExtGal_DM_limits} shows the limits derived by the analysis assuming the $\chi \rightarrow \tau^{+}\tau^{-}$ scenario. These results complement the limits from previous IceCube analyses and gamma-ray experiments. While PeV gamma-rays would be attenuated due to Galactic and extragalactic background light~\cite{ExtGal_bkg,Gal_bkg}, neutrinos are not subject to such an attenuation. Therefore, they are an excellent complementary tool to search for extragalactic heavy DM decay. 

\begin{figure}[bt]
\centering
    \includegraphics[width=0.44\linewidth]{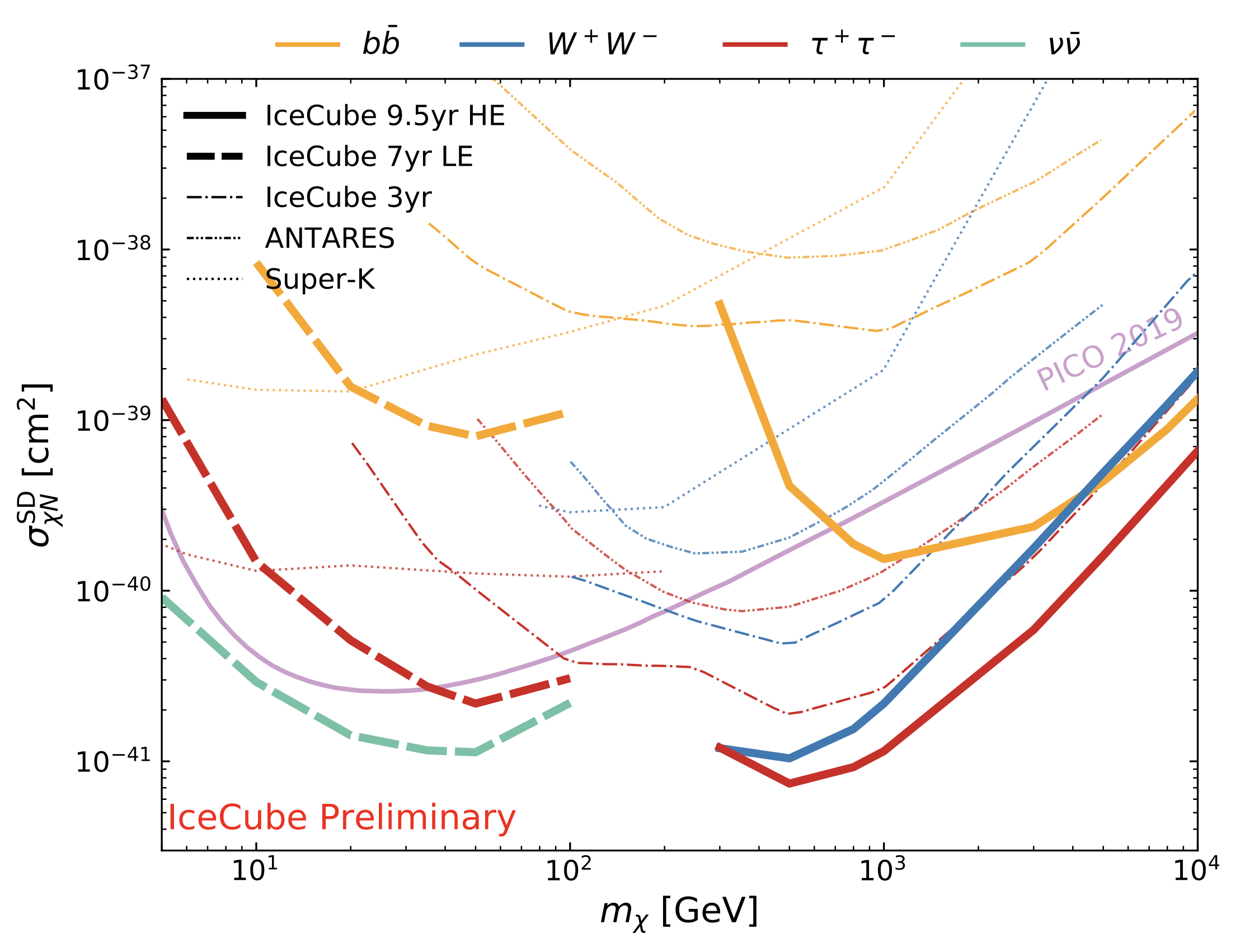}
    \includegraphics[width=0.44\linewidth]{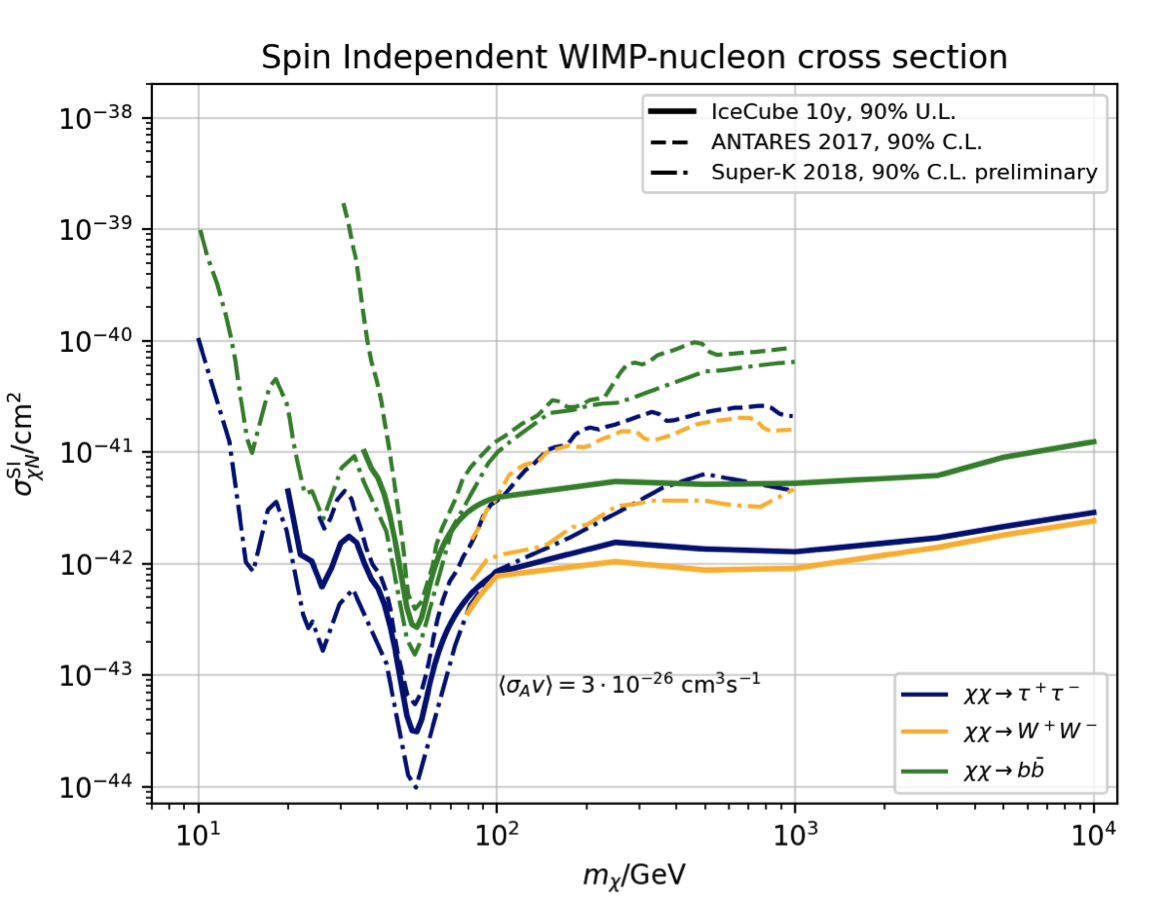}
    \caption{Limits on DM-nucleon scattering cross section. In the left panel, the thick lines, labeled "9.5 yr HE" and "7 yr LE", represent limits from our two recent solar DM searches. These limits were calculated assuming four different DM annihilation channels and are compared with previous neutrino experiments. In the right panel, the solid lines are limits from our latest Earth DM search, calculated for three different annihilation channels. They are compared with previous Super-K and ANTARES analyses. To calculate these limits the velocity-averaged DM annihilation cross section is assumed to be $3\times 10^{-26}$ cm$^{-2}$s$^{-1}$. These plots are taken from References~\protect\cite{NuDM2022,ICRC2023_Earth_DM}. }
    \label{fig:Solar_and_Earth_DM_limits}
\end{figure}

Neutrinos provide a unique way to probe DM-nucleon scattering. As the solar system travels through the Galactic DM halo, DM particles could scatter off nucleons in the Sun and lose enough kinetic energy to become gravitationally trapped in the solar core. The accumulated DM particles in the solar core could then self-annihilate to produce Standard Model particles. Among these products, only neutrinos can escape the Sun and be detectable at Earth. The DM density in the solar core depends on the capture rate, the self-annihilation rate, and the evaporation rate, which is negligible for DM masses above a few GeV. Given the age of the Sun, the capture and annihilation rates are expected to be in equilibrium~\cite{solar_WIMP_theory}. Thus, the neutrino flux from DM annihilation can be used to probe the DM-nucleon scattering cross section. DM particles are also expected to accumulate and annihilate in the Earth's core. However,  equilibrium between the capture and annihilation rates is not expected for the core of the Earth. Hence, an assumed annihilation rate is needed to constrain the scattering cross section using neutrinos from the Earth's core~\cite{Earth_DM}. Figure~\ref{fig:Solar_and_Earth_DM_limits} presents the limits obtained from recent IceCube searches for DM in the Sun and Earth~\cite{solar_DM_LE,solar_DM_HE,ICRC2023_Earth_DM}.    

\section{Conclusions}
We presented recent results from IceCube on diffuse astrophysical neutrino flux measurements, point source searches, and dark matter searches, and also discussed on-going studies. The diffuse flux measurements by the ESTES and global-fit analyses are consistent with a SPL spectrum hypothesis and agree with the previous IceCube measurements. Moreover, the global-fit analysis has achieved unprecedented precision in the flux measurement under the SPL hypothesis. Our recent point source searches hint at possible neutrino emission from the directions of NGC 4151 and CGCG 420-015, although the observed significance is below 3~$\sigma$. Multimessenger observations are expected to have sufficient sensitivity to test these sources hypotheses. Lastly, IceCube maintains an active program of dark matter searches and has derived competitive limits on the dark matter models. 

\section*{Acknowledgements}
The support and resources from the Center for High Performance Computing at the University of Utah are gratefully acknowledged.

\end{document}